\documentclass[12pt]{article}

\RequirePackage{amsthm,amsmath,amsfonts,amssymb}
\RequirePackage[colorlinks,citecolor=blue,urlcolor=blue]{hyperref}
\RequirePackage{graphicx}
\usepackage{multirow}
\usepackage{color}
\usepackage{mathrsfs}
\usepackage{caption,subcaption}
\usepackage[section]{placeins}
\usepackage{caption}
\usepackage{newclude}
\usepackage{algorithmicx}
\usepackage[ruled,vlined]{algorithm2e}

\def\T{T}

\def\Ebb{E}
\def\m{\mathcal{M}}
\def\real{R}
\def\exp{\mathrm{Exp}}
\def\sym{\mathrm{Sym}^+(m)}
\def\sy{\mathrm{Sym}(m)}
\def\mexp{\mathrm{exp}}

\def\log{\mathrm{Log}}
\def\mlog{\mathrm{log}}
\def\aj{a_j}
\def\bj{b_j}
\def\vec{\mathrm{vec}}
\def\vecs{\mathrm{vecs}}
\def\vecss{\mathrm{vecss}}
\def\h{h}

\newcommand{\RN}[1]{%
	\textup{\uppercase\expandafter{\romannumeral#1}}
}

\newtheorem{theorem}{Theorem}[section]

\newtheorem{corollary}[theorem]{Corollary}

\usepackage{hyperref}
\usepackage{amsmath}
\usepackage{graphicx}
\usepackage{enumerate}
\usepackage{natbib}
\usepackage{url} 

\newcommand{\blind}{0}

\begin{document}

	\def\spacingset#1{\renewcommand{\baselinestretch}%
		{#1}\small\normalsize} \spacingset{1}

		\if0\blind
	{
		\title{\bf Intrinsic Minimum Average Variance Estimation for Sufficient Dimension Reduction with Symmetric Positive\\
			Definite Matrices and Beyond}
		\author{Baiyu Chen\\
			School of Statistics, East China Normal University\\
			and \\
			Shuang Dai \\
			School of Statistics, East China Normal University\\
			and \\
			Zhou Yu \\
			School of Statistics, East China Normal University}
		\date{}
		\maketitle
	} \fi

	\if1\blind
	{
		\bigskip
		\bigskip
		\bigskip
		\begin{center}
			{\LARGE\bf Intrinsic Minimum Average Variance Estimation for Sufficient Dimension Reduction with Symmetric Positive\\
				 Definite Matrices and Beyond}
		\end{center}
		\medskip
	} \fi
	
	\bigskip
	\begin{abstract}
		In this paper, we target the problem of sufficient dimension reduction with symmetric positive definite matrices valued responses. We propose the intrinsic minimum average variance estimation method and the intrinsic outer product gradient method which fully exploit the geometric structure of the Riemannian manifold where responses lie. We present the algorithms for our newly developed methods under the log-Euclidean metric and the log-Cholesky metric. Each of the two metrics is linked to an abelian Lie group structure that transforms our model defined on a manifold into a Euclidean one. The proposed methods are then further extended to general Riemannian manifolds. We establish rigourous asymptotic results for the proposed estimators, including the rate of convergence and the asymptotic normality. We also develop a cross validation algorithm for the estimation of the structural dimension with theoretical guarantee Comprehensive simulation studies and an application to the New York taxi network data are performed to show the superiority of the proposed methods. 
	\end{abstract}
	
	\noindent%
	\vfill
	
	\newpage
	\spacingset{1.9} 
	\section{Introduction} \label{sec1}
	Undergoing accelerated developments for more than 20 years, sufficient dimension reduction (SDR) has now become a powerful tool in statistics, partly thanks to the increasing demand for techniques to deal with high-dimensional circumstances. Multiple classes of SDR methods have evolved themselves to be well-established for high-dimensional data analysis with Euclidean responses and predictors. Typical SDR tools include the inverse regression estimation methods (e.g., Sliced Inverse Regression (Li, 1991), Sliced Average Variance Estimation (Cook and Weisberg, 1991) and directional regression (Li and Wang, 2007)), the nonparametric method like the outer product of gradients (OPG) and the minimum average variance estimation (MAVE) method (Xia et al., 2002), and the semiparametric approach (Ma and Zhu, 2012, 2013, 2019).
	
	However, the prosperity of big data is accomplished by the abundance of non-Euclidean objects where traditional dimension reduction methods fail. For example, in an Alzheimer's Disease Neuroimaging Initiative (ANDI) study (Lin et al., 2022), subjects were invited to a medical center to get their brain images and assessment of their behavior abilities. Then a preprocessing protocol is applied to turn brain images into the average hippocampal diffusion tensors which are $3\times3$ symmetric positive definite (SPD) matrices characterizing diffusion of water molecules in tissues and conveying rich information about brain tissues. Finally researchers are faced with a data set $(Y_i,X_{1i},X_{2i},...,X_{pi})$ $(i=1,...,n)$ where the response $Y$ is a $3\times3$ SPD matrix and $X_1,...,X_p$ are predictors standardized to the interval $[0,1]$ representing the scores of each subject's memory, executive functioning, language ability and so on. Another example arises from the taxi services within a city. Researchers divide the city into several zones and take these zones as nodes in a network or graph. This graph is further weighted by the number of taxi pick-ups and drop-offs between zones in a time interval. Proper transformations can turn these graphs into SPD matrices describing the taxi movements in a city. After collecting potential predictors such as travel distance, fare amount, average daily temperature and total precipitation, one can analyze the relationship between the taxi movements and possible factors.
	
	In above examples, responses are non-Euclidean and lie in  $\mathrm{Sym}^+(m)$ which stands for a manifold consisting of $m\times m$ SPD matrices. When the dimension of prediction variable is large, sufficient dimension reduction is necessary to avoid the curse of dimensionality but unfortunately, traditional Euclidean methods cannot work for responses being SPD matrices. As a consequence, there has been a growing need to carry out SDR with SPD matrices as responses.
	
	Up to now there have been many works where traditional statistical methods in Euclidean spaces are generalized to manifolds or more general metric spaces such as local polynomial regression for SPD matrices (Yuan et al., 2012; Zhu et al., 2009; Cornea et al., 2016), Fr\'echet regression for random objects (Peterson and M\"uller, 2019a), intrinsic Riemannian functional principal component analysis and functional linear regression (Lin and Yao, 2019), additive model for SPD matrices (Lin et al., 2022), Fr\'echet sufficient dimension reduction for random objects (Ying and Yu, 2022; Zhang et al., 2021), intrinsic Wasserstein correlation analysis (Zhou et al., 2021), single index Fr\'echet regression (Bhattacharjee and M\"uller, 2021), autoregressive optimal transport model (Zhu and M\"uller, 2021) and so on. 
	
	Among these works, two recent papers are related to non-Euclidean SDR. Ying and Yu (2022) extended the traditional SIR model to the case where the predictors are Euclidean while the response takes values in a metric space. They borrowed strength from the martingale difference divergence to avoid the estimation of $\Ebb(X\mid Y)$ and to absorb information in $Y$ by including the distance function in the metric space. The work of Zhang et al. (2021) turned almost all existing Euclidean SDR methods into ones for Euclidean $X$ and metric space-valued $Y$, which is very comprehensive and flexible.
	
	In their proposal, the random object $Y$ is first mapped into a real-valued random variable and then classic SDR methods can be applied to the transformed data. However, when the response lie in a manifold, even though the two methods aforementioned can be performed, they fail to fully exploit the intrinsic geometry of the manifold and thus some information contained in the response is inevitably lost.
	
	In this paper, we consider the dimension reduction of the conditional mean (Cook and Li, 2002) with SPD matrices. The basic problem is to find a lower dimensional predictor $B_0^\T X$ such that   
	\begin{equation}\label{CMS}
		\Ebb(Y\mid X)= \Ebb(Y\mid B_0^\T X),
	\end{equation}
 where $Y\in \text{Sym}^+(m)$ and $B_0$ is a $p\times d$ matrix. To fully incorporate the information in the $\mathrm{Sym}^+(m)$-valued response, we generalize the state-of-the-art sufficient mean dimension reduction method MAVE and OPG for the estimation of the column space spanned by $B_0$. The basic idea of our method also stems from the local polynomial regression (ILPR) for SPD matrices introduced by Yuan et al. (2012), which replaced the square distance by the geodesic distance on $\mathrm{Sym}^+(m)$ and performed Taylor expansion after parallel transport to estimate an intrinsic conditional expectation of an SPD matrix response, given a covariate vector $X$. Yuan et al. (2012) only considered the case where $X$ is a scalar. We  in this paper take a step forward to handle the high-dimensional $X$. We call our method intrinsic MAVE and intrinsic OPG since $\mathrm{Sym}^+(m)$ cannot be isometrically embedded into a Euclidean space and we deal with it in a totally intrinsic way.
	
The rest of this paper is organized as follows. Some preliminaries on manifolds are introduced in Section 2. Then we introduce our intrinsic dimension reduction proposals and algorithms with SPD matrices in Section 3 and Section 4. Our proposed methods for SPD matrices are extended to general manifolds in Section 5. Asymptotic results, including the rate of convergence and asymptotic normality are established in Section 6. A cross validation procedure to determine the structural dimension is presented in Section 7. Simulation studies are illustrated in Section 8 and a real data application is carried out in Section 9. Section 10 concludes this paper. Additional simulation results and proofs for theorems can be found in the supplementary material.
	
	\section{Preliminaries on Manifolds}
	We first introduce some basic notions for Riemannian manifolds and Lie groups (Tu, 2011; Lang, 1999). Let $\m$ be a simply connected and smooth manifold and $p\in\m$. For a small scalar $\delta>0$, let $c(t)$ be a continuously differential map from $(-\delta,\delta)$ to $\m$ passing through $c(0)=p$. A tangent vector at $p$ is the derivative of the  curve $c(t)$ at $t=0$. All such tangent vectors at $p$ form a vector space named the tangent space at $p$, which is denoted by $T_p\m$. Each tangent space $T_p\m$ can be endowed with an inner product $\langle\cdot,\cdot\rangle_p$ that varies smoothly with $p$. The inner products $\{\langle\cdot,\cdot\rangle_p:p\in\m\}$ are collectively denoted by $\langle\cdot,\cdot\rangle$, which is referred to as the Riemannian metric of $\m$. With a Riemannian metric, we can define a distance $d(\cdot,\cdot)$ on $\m$ that turns $\m$ into a metric space. The length of a continuously differentiable curve $c(t):[t_0,t_1]\rightarrow\m$ is calculated as $\int_{t_0}^{t_1}\langle c'(t),c'(t)\rangle_{c(t)}^{1/2}\mathrm{d}t$, where $c'(t)$ is the derivative of $c(t)$. And $d(p,q)$ is the infimum of the length over all continuously differentiable curves joining $p$ and $q$.
	
	A geodesic $\gamma$ is a curve defined on $[0,\infty)$ such that for each $t\in[0,\infty)$, $\gamma([t,t+\epsilon])$ is the shortest path connecting $\gamma(t)$ and $\gamma(t+\epsilon)$ for sufficiently small $\epsilon>0$. The Riemannian exponential map $\exp_p$ at $p\in\m$ is a function mapping $T_p\m$ into $\m$ and defined by $\exp_p(u)=\gamma(1)$ with $\gamma(0)=p$ and $\gamma'(0)=u\in T_p\m$. The inverse of $\exp_p$, if exists, denoted by $\log_p$ and called the Riemannian logarithm map at $p$, can be defined as $\log_pq=u$ for $q\in\m$ such that $\exp_pu=q$. 
	
	A vector field $U$ is a function defined on $\m$ such that $U(p)\in T_p\m$. Given a curve $\gamma(t)$ on $\m$, $t\in I$ for a real interval $I$, a vector field along $\gamma$ is a smooth map defined on $I$ such that $U(t)\in T_{\gamma(t)}\m$. We say $U$ is parallel along $\gamma$ if $\triangledown_{\gamma'(t)}U=0$ for all $t\in I$ where $\triangledown$ is the Levi-Civita connection on $\m$. In this paper we only focus on parallel vector fields along geodesics. Let $\gamma:[0,1]\rightarrow\m$ be a geodesic connecting $p$ and $q$, and $U$ is a parallel vector field along $\gamma$ such that $U(0)=u$ and $U(1)=v$. Then the parallel transport of $u$ along $\gamma$ is denoted as $\phi_p(u)=v$.

	When $(\m,\oplus)$ is a group and the group operation $\oplus$ and its inverse are both smooth, $(\m,\oplus)$ is called a Lie group. The tangent space at the identity element $e$ is called a Lie algebra denoted by $\mathfrak{g}$. It consists of left-invariant vector fields $U$ which satisfies $U(p\oplus q)=(DL_p)(U(q))$, where $L_p:q\rightarrow p\oplus q$ is the left translation at $p$ and $DL_p$ is the differential of $L_p$. A Riemannian metric $\langle\cdot,\cdot\rangle$ is called left-invariant if $\langle u,v\rangle_q=\langle DL_p(u),DL_p(v)\rangle_{p\oplus q}$ for all $p,q\in\m$ and $u,v\in T_q\m$. Right invariance can be defined similarly. A metric is bi-invariant if it is both left-invariant and right-invariant. The Lie exponential map, denoted by $\mathfrak{exp}$ is defined by $\mathfrak{exp}(u)=\gamma(1)$ where $\gamma:\real\rightarrow\m$ is the unique one-parameter subgroup such that $\gamma'(0)=u\in\mathfrak{g}$. Its inverse, if exists, is denoted by $\mathfrak{log}$. Please make a distinction between the Riemannian exponential map “$\exp$”, the Lie exponential map “$\mathfrak{exp}$” and the common matrix exponential operation “$\mexp$” which appear frequently in later sections. When $\langle\cdot,\cdot\rangle$ is bi-invariant, then $\mathfrak{exp}$ coincides with $\exp_e$.

\section{Intrinsic MAVE and OPG for SPD Matrices}

The classic MAVE in a Euclidean space adopts the following regression-type model for conditional mean dimension reduction:
\begin{equation}\label{model}
	Y=g(B_0^\T X)+\varepsilon,
\end{equation}
where $Y$ and $X$ are respectively $\real$-valued and $\real^p$-valued random variables, $g$ is an unknown smooth link function, $B_0=(\beta_1,...,\beta_d)$ is a $p\times d$ orthogonal matrix ($B_0^\T B_0=I_{d\times d}$) with $d<p$ and $E(\varepsilon\mid X)=0$ almost surely. MAVE aims to estimate $B_0$ as $B_0^\T X$ captures all information about $Y$ provided by $X$.

MAVE targets $B_0$ by solving
\begin{equation}\nonumber
	\mathop{\min}_{B:B^{\T}B=I}\Ebb\{Y-\Ebb(Y\mid B^\T X)\}^2,
\end{equation}
which is equivalent to
\begin{equation}\nonumber
	\min_{B:B^{\T}B=I}\Ebb\left(\Ebb\left[\left\{Y-\Ebb(Y\mid B^TX)\right\}^2\mid B^\T X\right]\right).
\end{equation}

Suppose $(Y_i,X_i)$ $(i=1,...,n)$ is a sample from $(Y,X)$. Following the idea of local linear regression, the above formula can be approximated by
\begin{equation}\label{start}
	\min_{B:B^{\T}B=I}\sum_{j=1}^{n}\sum_{i=1}^{n}w_{ij}\left\{Y_{i}-\Ebb(Y_i\mid B^\T X_i)\right\}^2,
\end{equation}
which can be further approximated by 
\begin{equation}\label{target}
	\min_{B:B^{\T}B=I\atop a_{j},b_{j}}\sum_{j=1}^{n}\sum_{i=1}^{n}w_{ij}\left[Y_{i}-\left\{a_{j}+b_{j}^{\T}B^{\T}(X_{i}-X_{j})\right\}\right]^2,
\end{equation}
where $w_{ij}=K_h(X_i-X_j)/\sum_{i=1}^nK_h(X_i-X_j)$ and for $u\in \real^p$, $K_h(u)=K(u/h)/h^p$ with $K(\cdot)$ being the kernel function and $h\in\real$ being the bandwidth. Optimizing (\ref{target}) gives the estimation of $B_0$.

When it comes to the manifold case where $X\in\real^p$ but $Y\in\sym$, model (\ref{model}) should be modified. In this case, $g:\real^d\rightarrow\sym$ is a link function and $g(B_0^\T X),\varepsilon\in\sym$. In order to ensure that $Y\in\sym$, we assume a group structure on $\sym$ with the group operator $\oplus$, and replace  $+$ by $\oplus$. To make our model more flexible, we further assume that $(\sym,\oplus)$ is a commutative group (abelian group).

Let $(\sym,\oplus)$ be an abelian group endowed with a Riemannian metric $\langle\cdot,\cdot\rangle$. Let $g:\real^d\rightarrow\sym$ be the link function and $\varepsilon\in\sym$ be the random noise whose Fr\'echet mean corresponds to the group identity element. Then conditional mean sufficient dimension reduction with $X\in\real^p$ but $Y\in\sym$ can be formulated as
\begin{equation}\label{ourmodel}
	Y=g(B_0^\T X)\oplus\varepsilon.
\end{equation}

We first figure out the definition of conditional expectation $\mathbb{E}(Y\mid B^\T X)$ when $Y$ is an SPD matrix. According to Yuan et al. (2012), the intrinsic conditional expectation  of $Y$ at $B^\T X=B^\T x$ is defined as a SPD matrix $D(B^\T x)\in\sym$ such that
\begin{equation}\nonumber
	\Ebb\left\{\log_{D(B^\T x)}Y\mid B^\T x\right\}=O_m,
\end{equation}
where $O_m$ is an $m\times m$ matrix with all elements 0 and the expectation is taken in a component-wise way. From now on we use $D(B^\T x)$ instead of $\Ebb(Y\mid B^\T x)$.

Starting from (\ref{start}), we replace the square distance by the squared geodesic distance $d^2(\cdot,\cdot)$ on the manifold and rewrite (\ref{start}) as
\begin{equation}\label{ready}
	\min_{B:B^{\T}B=I}\sum_{j=1}^{n}\sum_{i=1}^{n}w_{ij} d^2\left\{Y_{i},D(B^\T X_i)\right\}.
\end{equation}

Next we want to similarly expand $D(B^\T x)$ at $B^\T X_j$. However, $D(B^\T x)$ is in the curved space and Taylor expansion is infeasible. Instead we apply the Riemannian logarithm map to transform $D(B^\T x)$ to $\log_{D(B^\T X_j)}D(B^\T x)\in T_{D(B^\T X_j)}\sym$. Since $\log_{D(B^\T X_j)}D(B^\T x)$ for different $X_j$ are in different tangent spaces, these tangent vectors are transported from $T_{D(B^\T X_j)}\sym$ to a same tangent space $T_{I_m}\sym$ using parallel transport given by:
\begin{equation}\nonumber
	\phi_{D(B^\T X_j)}: T_{D(B^\T X_j)}\sym\rightarrow T_{I_m}\sym.
\end{equation}

Thus $f(B^\T x)=\phi_{D(B^\T X_j)}\log_{D(B^\T X_j)}D(B^\T x)$ is a function in a vector space and can be expanded at $B^TX_j$ using Taylor series expansion. Considering $f(B^\T x)$ is an $m\times m$ symmetric matrix and $B^\T X_j$ is a $d\times1$ vector, we differentiate each component of $f(B^\T x)$ with respect to $B^\T X_j$ and this leads to
\begin{equation}\label{appro}
	\begin{split}
		\log_{D(B^\T X_j)}D(B^\T x)=&\phi_{D(B^\T X_j)}^{-1}\{f(B^\T x)\}\\
		\approx&\phi_{D(B^\T X_j)}^{-1}\left[b_j I_m\otimes \left\{B^\T(x-X_j)\right\}\right],
	\end{split}
\end{equation} 
which gives
\begin{equation}\nonumber
	D(B^Tx)\approx\exp_{D(B^\T X_j)}\left(\phi_{D(B^\T X_j)}^{-1}\left[b_jI_m\otimes \left\{B^\T(x-X_j)\right\}\right]\right),
\end{equation} 
where only up to first order approximation is considered,  $\phi_{D(B^\T X_j)}^{-1}$ is the inverse map of $\phi_{D(B^\T X_j)}$ and $\otimes$ is the Kronecker product.

In (\ref{appro}), both $D(B^\T X_j)$ and $b_j$ are parameters to estimate: $D(B^\T X_j)\in\sy$ serving as the 0-order approximation in Taylor expansion, $b_j$ being the derivative matrix in the first order term and possessing the structure
\begin{equation}\label{y1}
	b_j=\left(
	\begin{array}{cccc}
		c_{11}^{\T}(X_{j})&c_{12}^{\T}(X_{j})&\cdots&c_{1m}^{\T}(X_{j})\\
		c_{21}^{\T}(X_{j})&c_{22}^{\T}(X_{j})&\cdots&c_{2m}^{\T}(X_{j})\\
		\vdots&\vdots& &\vdots\\
		c_{m1}^{\T}(X_{j})&c_{m2}^{\T}(X_{j})&\cdots&c_{mm}^{\T}(X_{j})
	\end{array}
	\right)_{m\times md}\quad (j=1,...,n),
\end{equation}
where $c_{kl}(X_j)=c_{lk}(X_j)\in\real^d$ $(k,l=1,...,m)$. The $X_{j}$ in parentheses indicate that $b_j$ is related to $X_{j}$. We use $a_j$ to denote $D(B^\T X_j)$ for simplicity here and hereafter.

Now we introduce three operators in matrix algebra. ``$\vec(\cdot)$" is the common matrix vec operator that vectorize an $m\times n$ matrix by column into an $mn\times1$ vector. For an $m\times m$ symmetric matrix $A=(a_{ij})$, define $\vecs(A)=(a_{11},a_{21},a_{22},...,a_{m1},...,a_{mm})^\T$. That is, ``$\vecs(\cdot)$" vectorize the lower triangle part of a symmetric matrix by row. For $b_j$ in (\ref{y1}), define $\vecss(b_j)=(c_{11}^{\T}(X_{j}),c_{21}^{\T}(X_{j}),c_{22}^{\T}(X_{j}),...,c_{m1}^{\T}(X_{j}),...,c_{mm}^{\T}(X_{j}))^\T$. We will frequently use $\vec(B)$, $\vecs(a_j)$ and $\vecss(b_j)$ in subsequent sections.

Finally combining (\ref{appro}) with (\ref{ready}), we arrive at what we call the intrinsic MAVE method (iMAVE):
\begin{equation}\label{imave}
	\min_{B:B^{\T}B=I\atop \aj,\bj}\sum_{j=1}^{n}\sum_{i=1}^{n}w_{ij} d^2\left\{Y_{i}\, , \, \exp_{\aj}\left( \phi_{\aj}^{-1}\left[ \bj I_m\otimes \left\{B^{\T}(X_{i}-X_{j})\right\}\right] \right)\right\},
\end{equation}
where $a_j$ is  $m\times m$ and $b_j$ is $m\times md$.

The only difference between the classic OPG and the classic MAVE is the absence of $B$ in the former. So the intrinsic OPG method (iOPG) can be formulated immediately as 
\begin{equation}\label{iopg}
	\min_{\aj,\bj}\sum_{j=1}^{n}\sum_{i=1}^{n}w_{ij} d^2\left(Y_{i}\, , \, \exp_{\aj}\left[ \phi_{\aj}^{-1}\left\{ \bj I_m\otimes (X_{i}-X_{j})\right\} \right]\right),
\end{equation}
where the size of $b_j$ here is $m\times mp$. 

Only the Riemannian metric needs specifying to solve (\ref{imave}) and (\ref{iopg}). Actually we do not require (\ref{ourmodel}) to be true since the procedure of deriving iMAVE and iOPG has nothing to do with the group structure on $\sym$. We only assume (\ref{ourmodel}) when performing the theoretical analysis. Thus (\ref{imave}) and (\ref{iopg}) are flexible SDR methods but the choice of the metric affects the complexity of optimization.

\section{Algorithms under the Log-Euclidean Metric}
The log-Euclidean metric is proposed by Arsigny et al. (2007). The key observation is that: $\sym$ is diffeomorphic to its tangent space at the identity matrix, $\sy$. To be specific, $\mexp:\sy\rightarrow\sym$ and its inverse $\mlog$ are both smooth and they are diffeomorphisms.

Let $S_1,S_2\in\sym$. Define an operation $\oplus$ by
\begin{equation}\label{groupop}
	S_1\oplus S_2=\mexp\{\mlog(S_1)+\mlog(S_2)\}.
\end{equation}  

Then $(\sym,\oplus)$ is an abelian Lie group. The identity element is the identity matrix. Moreover, the Lie group exponential map $\mathfrak{exp}$ is just the matrix exponential $\mexp$. That is, the matrix logarithm $\mlog$ maps every SPD matrix in $\sym$ to the tangent space $T_{I_m}\sym$. Based on this fact, we may get the expression of iMAVE under the log-Euclidean metric in a simpler way.

We start from (\ref{ready}). Under the log-Euclidean metric, the geodesic distance $d(S_1,S_2) = ||\mlog S_1-\mlog S_2||_F$. Here $||\cdot||_F$ is the Frobenius norm. So
\begin{equation}\nonumber
	d\left\{Y_{i},D(B^\T X_i)\right\}=||\mlog\{D(B^\T X_i)\}- \mlog Y_{i}||_F.
\end{equation}

Since $\mlog\{D(B^\T X_i)\}$ and $\mlog Y_{i}$ are both in $T_{I_m}\sym$, no parallel transportation is needed. Directly expand $\mlog\{D(B^\T X_i)\}$ at $B^\T X_j$, we get iMAVE under the log-Euclidean metric:
\begin{equation}\label{eumave}
	\min_{B:B^{\T}B=I\atop \aj,\bj}\sum_{j=1}^{n}\sum_{i=1}^{n}w_{ij}||\aj+\bj I_m\otimes\{B^{\T}(X_{i}-X_{j})\}-\mlog Y_i   ||_F^2,
\end{equation}
and similarly iOPG under the log-Euclidean metric:
\begin{equation}\label{euopg}
	\min_{\aj,\bj}\sum_{j=1}^{n}\sum_{i=1}^{n}w_{ij}||\aj+\bj I_m\otimes(X_{i}-X_{j})-\mlog Y_i   ||_F^2.
\end{equation}

Models (\ref{eumave}) and (\ref{euopg}) are optimized similarly to Xia et al. (2002) or Xia (2007). The main difference here is differentiating a matrix-valued function w.r.t. a vector. We in the following sketch out the algorithms. First some notations are introduced.

Write $q=m(m+1)/2$ and let
\begin{equation}\nonumber
	\begin{split}
		& w_{ij}=\frac{K_h(B^\T(X_i-X_j))}{\sum_{i=1}^nK_h(B^\T(X_i-X_j))},\quad \alpha_j=\left(
		\begin{array}{c}
			\vecs(\aj)\\
			\vecss(\bj)
		\end{array}
		\right),\\
		&\chi_i(X_j)=\Big(I_q,I_q\otimes (X_{i}-X_{j})^\T\Big)^\T, \quad\chi_i(B^\T X_j)=\Big(I_q,I_q\otimes \big((X_{i}-X_{j})^\T B\big)\Big)^\T,\\
		&A_{ij}=\Big(c_{11}(X_j),c_{21}(X_j),c_{22}(X_j),...,c_{m1}(X_j),...,c_{mm}(X_j)\Big)\otimes(X_i-X_j),
	\end{split}
\end{equation}
where $c_{kl}(X_j)$ $(1\leq l\leq k\leq m)$ are from (\ref{y1}).

Now we are ready for the algorithms.

\begin{algorithm}
	\caption{iMAVE under the log-Euclidean metric.}
	\label{algorithmmave}
	
	\textbf{Step 1.} Marginally standardize $X_1,...,X_n$ when necessary. Set the bandwidth $h_0=c_0n^{-1/(p_0+6)}$, where $c_0=2.34$ and $p_0=\max(p,3)$. Let $\hat{B}_{(0)}$ be an initial estimator. Set $t=1$.
	
	\textbf{Step 2.} Compute
	\begin{equation}\nonumber
		\begin{split}
			\hat{\alpha}_j^{(t)}=&\Big\{\sum_{i=1}^{n}w_{ij}^{(t-1)}\chi_{i}(\hat{B}_{(t-1)}^\T X_{j})\chi_{i}(\hat{B}_{(t-1)}^\T X_{j})^\T\Big\}^{-1}\\
			&\times\sum_{i=1}^{n}w_{ij}^{(t-1)}\chi_{i}(\hat{B}_{(t-1)}^\T X_{j})\vecs(\mlog Y_{i})\quad	(j=1,...,n).
		\end{split}
	\end{equation}
	Read off $\vecs(\hat{a}_j^{(t)})$ and $\vecss(\hat{b}_j^{(t)})$ respectively from the first $q$ and the remaining $qd$ components of $\hat{\alpha}_j^{(t)}$.
	
	\textbf{Step 3.} Compute
	\begin{equation}\nonumber
		\vec(\hat{B}_{(t)})=\Big\{\sum_{j=1}^{n}\sum_{i=1}^{n}w_{ij}^{(t-1)}A_{ij}^{(t)} (A_{ij}^{(t)})^\T\Big\}^{-1}\sum_{j=1}^{n}\sum_{i=1}^{n}w_{ij}^{(t-1)}A_{ij}^{(t)}\vecs(\mlog Y_{i}-\hat{a}_j^{(t)}).
	\end{equation}
	
	\textbf{Step 4.} If $t<30$, reset $h_{t+1}=\max(r_nh_{t},c_0n^{-1/(d+4)})$, where $r_n=n^{-1/2(p_0+6))}$. Set $t=t+1$ and go back to step 2. Otherwise, get the iMAVE estimator $\hat{B}_{(t)}$.
\end{algorithm}

Similarly the algorithm of iOPG is shown in Algorithm \ref{algorithmopg}. Usually the result of OPG can be used as the initial value of $B$ in MAVE.
\begin{algorithm}
	\caption{iOPG under the log-Euclidean metric.}
	\label{algorithmopg}
	
	\textbf{Step1.} Marginally standardize $X_1,...,X_n$ when necessary. Set the bandwidth $h_0=c_0n^{-1/(p_0+6)}$, where $c_0=2.34$ and $p_0=\max(p,3)$. Set $\hat{B}_{(0)}=I_p$. Set iteration time $t=1$.
	
	\textbf{Step2.} Compute
	\begin{equation}\nonumber
		\begin{split}
			\hat{\alpha}_j^{(t)}=\left\{\sum_{i=1}^{n}w_{ij}^{(t-1)}\chi_{i}(X_{j})\chi_{i}(X_{j})^\T\right\}^{-1}\sum_{i=1}^{n}w_{ij}^{(t-1)}\chi_{i}(X_{j})\vecs(\mlog Y_i)\quad(j=1,...,n).		\end{split}
	\end{equation}
	
	Read off $\vecss(\hat{b}_j^{(t)})$ from the last $qd$ components of $\hat{\alpha}_j^{(t)}$.
	
	\textbf{Step3.} Recover $\hat{b}_j^{(t)}, j=1,...,n$ from $\vecss(\hat{b}_j^{(t)})$ in step 2 as
	\begin{equation}\nonumber
		\hat{b}_j^{(t)}=\left(
		\begin{array}{cccc}
			c_{11}^{\T}& & &\\
			c_{21}^{\T}&c_{22}^{\T}& &\\
			\vdots&\vdots& \ddots&\\
			c_{m1}^{\T}&c_{m2}^{\T}&\cdots&c_{mm}^{\T}
		\end{array}
		\right)\quad (j=1,...,n),
	\end{equation}
	with the symmetric part omitted. Rearrange the lower triangle part of $\hat{b}_j^{(t)}$ to get
	\begin{equation}\nonumber
		\hat{\beta}_j^{(t)}=(c_{11},c_{21},c_{22},...,c_{m1},...,c_{mm})^\T\in\real^{q\times p},j=1,...,n.
	\end{equation}
	
	\textbf{Step 4.} Compute
	\begin{equation}\nonumber
		\hat{\Lambda}^{(t)}=\frac{1}{n}\sum_{j=1}^{n}(\hat{\beta}_j^{(t)})^\T\hat{\beta}_j^{(t)}.
	\end{equation}
	
	Perform eigen-decomposition for $\hat{\Lambda}^{(t)}$ and get the $d$ eigenvectors $\hat{v}_1,...,\hat{v}_d$ corresponding to its largest $d$ eigenvalues. Let $\hat{B}_{(t)}=(\hat{v}_1,...,\hat{v}_d)$.
	
	\textbf{Step 5.} If $t<30$, reset $h_{t+1}=\max(r_nh_{t},c_0n^{-1/(d+4)})$, where $r_n=n^{-1/2(p_0+6))}$. Set $t=t+1$ and go back to step 2. Otherwise, get the iOPG estimator $\hat{B}_{(t)}$.
\end{algorithm}

When $\sym$ is endowed the log-Cholesky metric, methods can be derived similarly as the key point is under the log-Cholesky metric, the geodesic distance between $S_1, S_2\in\sym$ is $d(S_1,S_2)=||\mathrm{chol}(L_1)-\mathrm{chol}(L_2)||_F$. Here $L_1,L_2$ are Cholesky factors of $S_1,S_2$ (Lin, 2019) and $\mathrm{chol}(L)=\lfloor L\rfloor +\mlog\mathbb{D}(L)$ where $\lfloor L\rfloor$ is the strict lower triangle part of $L$ and $\mathbb{D}(L)$ the diagonal part of $L$. For any $S\in\sym$ and its Cholesky factor $L$, $\mathrm{chol}(L)$ lies in a fixed vector space. Substituting $\mlog(\cdot)$ in the log-Euclidean case for $\mathrm{chol}(\cdot)$ and keeping other things unchanged, we get iMAVE, iOPG under the log-Cholesky metric and  details are omitted.
	
	\section{Extension to General Riemannian Manifolds}
	According to the lemma S1 in Lin (2022), if ($\m$,$\oplus$) is an abelian Lie group endowed with a bi-invariant metric $\langle\cdot,\cdot\rangle$ that turns $\m$ into a Hadamard manifold, for any $y,z,u,v\in\m$, $\log_y(y\oplus z)=\phi_{e,y}(\mathfrak{log}z)$, $\mathfrak{log}(u\oplus v)=\mathfrak{log}u+\mathfrak{log}v$. Here $e$ is the identity element of the group. Endowing $\sym$ with the log-Euclidean metric or the log-Cholesky metric can meet the conditions. So applying above equations to $Y=\mu\oplus g(B_0^\T X)\oplus\varepsilon$ which is equivalent to our model (\ref{ourmodel}) with $\mu$ denoting the Fr\'echet mean of $Y$, we have $\log_\mu Y=\phi_{e,\mu}\mathfrak{log}g(B_0^\T X)+\phi_{e,\mu}\mathfrak{log}\varepsilon$. This model can be rewritten as
	\begin{equation}\label{eucmodel}
		\log_\mu Y=h(B_0^\T X)+\zeta,
	\end{equation}
	where $h(\cdot)=\phi_{e,\mu}\mathfrak{log}g(\cdot)$ and $\zeta=\phi_{e,\mu}\mathfrak{log}\varepsilon$. The model (\ref{eucmodel}) is completely a Euclidean one since $h:R^d\rightarrow\sy$ is a vector-valued function defined in $R^p$, which brings convenience for the theoretical analysis of iMAVE and iOPG. However if the chosen metric cannot turn $\sym$ into an abelian group with a bi-invariant metric, neither (\ref{ourmodel}) nor (\ref{eucmodel}) holds. In this case, we can directly assume model (\ref{eucmodel}) for other metrics and furthermore general Riemannian manifolds.
	
	Let $X\in R^p$ and $Y\in\m$ where $(\m,\langle\cdot,\cdot\rangle)$ is a general Riemannian manifold. We assume the relationship between $X$ and $Y$ can be described by (\ref{eucmodel}). We still aim at estimating $B_0$ and the estimating procedure is just the MAVE and OPG with multivariate response developed by Zhang (2021), which can also be derived by slightly modifying our proposed algorithms in Section 4.
	
	For general Riemannian manifolds whose sectional curvature is positive, the Fr\'echet mean may not exist and therefore additional conditions are needed for (\ref{eucmodel}). We assume
	\begin{itemize}
	\item[(A1)] The minimizer of the Fr\'echet function $F(\cdot)=Ed^2(\cdot,Y)$ exists and is unique.
    \end{itemize}
	This is automatically satisfied when $\m$ is $\sym$ equipped with either the log-Euclidean metric or the log-Cholesky metric. 
	
	For a subset $A$ of $\m$, $A^\epsilon$ denotes the set $\cup_{p\in A}B(p;\epsilon)$ where $B(p;\epsilon)$ is the ball with center $p$ and radius $\epsilon$ in $\m$. We use $\mathrm{Im}^{-\epsilon}(\exp_{\mu})$ to denote the set $\m\setminus\{\m\setminus\mathrm{Im}(\exp_{\mu})\}^\epsilon$. In order to define $\log_{\hat{\mu}}Y_i$ at least with a dominant probability for a large sample, we assume
	\begin{itemize}
	\item[(A2)] There is some constant $\epsilon_0>0$ such that $\mathrm{pr}\{Y\in\mathrm{Im}^{-\epsilon}(\exp_{\mu})\}$=1.
	\end{itemize}
	The condition (A2) is only needed when $\m$ is not a Hadamard manifold. If (A1) and (A2) are satisfied, (\ref{eucmodel}) is well defined.
	
	\section{Asymptotic Results}
	We first establish the consistency and asymptotic normality of the iMAVE and iOPG estimators under the general manifolds case in model (\ref{eucmodel}) and the results of $\sym$ endowed with either the log-Euclidean metric or the log-Cholesky metric is given as corollaries. We consider a manifold $\m$ that satisfied one of the following conditions:
	\begin{itemize}
	\item[(M1)] $\m$ is a finite-dimensional Hadamard manifold having sectional curvature bounded from below by $\mathfrak{c}_0<0$.
	\item[(M2)] $\m$ is a complete compact Riemannian manifold.
	\end{itemize}
	An example satisfying (M1) is $\sym$ endowed with the log-Euclidean metric, the log-Cholesky metric
	or the affine-invariant metric while the unit sphere serves as an example satisfying (M2).
	
	We have to treat $\phi\log_{\hat{\mu}}Y_i-\log_\mu Y_i$ during our proof where $\phi$ is short for $\phi_{\hat{\mu},\mu}$. The method in Lin and Yao (2019) is applied here to write $\phi\log_{\hat{\mu}}Y_i-\log_\mu Y_i$ as $\{-H_i(\mu)+\Delta_i(\hat{\mu})\}\log_\mu\hat{\mu}$ and the asymptotic normality of $\log_\mu\hat{\mu}$ helps us control the discrepancy between $\log_{\hat{\mu}}Y_i$ and $\log_\mu Y_i$. Above $\Delta_i(\hat{\mu})=o_P(1)$ and $H_i(y)=-(\triangledown Z_i)(y)$, acting on vector fields $U,V$ by $\langle H_iU,V\rangle(y)=\langle -\triangledown_U Z_i,V\rangle(y)=\mathrm{Hess}_y\{d^2(y,Y_i)/2\}(U,V)$. Here $Z_i$ is a vector field with $Z_i(y)=\log_yY_i$ and ``$\mathrm{Hess}$'' denotes the Hessian matrix (Kendall and Le, 2011). To make above reasoning valid, following conditions are needed.
	\begin{itemize}
	\item[(A3)] $\m$ satisfies at least one of the conditions (M1) and (M2).
	
	\item[(A4)] For all $y\in \m$, $E\{d^2(y,Y)\}<\infty$.
	
	\item[(A5)] For some constant $\mathfrak{c}_1>0$, $F(y)-F(\mu)\geq \mathfrak{c}_1d^2(y,\mu)$ when $d(y,\mu)$ is sufficiently small.
	
	\item[(A6)] $\lambda_{\mathrm{min}}\{E(H_t)\}>0$ where $\lambda_{\mathrm{min}}(\cdot)$ is the smallest eigenvalue of an operator or a matrix.
	\end{itemize}
	Conditions (A3)-(A6) are standard assumptions also made by Lin (2022), Kedall and Le (2011) and Lin and Yao (2019). (A4) is analogous to the moment condition in the Euclidean case. (A5) is satisfied for Hadamard manifolds with $c_2=1$ according to the lemma S.7 of Lin and M\"{u}ller (2021). (A6) is made to ensure $H_i$ is invertible.
	
	We need additional conditions that are standard in the literature on MAVE and OPG methods such as Xia et al. (2002) and Xia (2007). 
	
	Some notations are listed here. Suppose the dimension of $\m$ is $s$ and thus terms in (\ref{eucmodel}) are $s$-dimensional vectors. Let $h_{k}(B_0^\T X)$ ($k=1,...,s$) denote the $k$th component of $h(B_0^\T X)$ and $\zeta_k$ are defined similarly. Let $\mu_B(u)=E(X\mid B^\T X=u)$, $w_B(u)=E(XX^\T\mid B^\T X=u)$, $v_B(u)=\mu_B(B^\T u)-u$, and $\bar{w}_B(u)=w_B(B^\T u)-\mu_B(B^\T u)\mu_B^\T(B^\T u)$, which will be frequently encountered in proofs.  For any square matrix $A$, $A^{-1}$ and $A^+$ denote the inverse (if it exists) and the Moore-Penrose inverse matrix.
	\begin{itemize}
	\item[(B1)] For $k=1,...,s$, $h_{k}(\cdot)$ has bounded, continuous third derivatives and $E(\zeta_k\mid X)=0$.
	
	\item[(B2)] The density function $f(x)$ of $X$ has bounded second order derivatives on $\real^p$ and is bounded away from 0 in a neighborhood around 0; $\Ebb|X|^r<\infty$ for some $r>8$; the functions $\mu_B(u)$ and $w_B(u)$ have bounded derivatives with respect to $u$ and $B$ for $B\in\{|B-B_0|<\delta\}$ for some $\delta>0$.
	
	\item[(B3)] For every component $y_k$ ($k=1,...,s$) in $\mlog Y$, the density function $f_{y_k}$ has bounded derivative and is bounded away from 0 on a  compact support; the conditional density functions $f_{y_k\mid X}(y\mid x)$ and $f_{y_k\mid B^\T X}(y\mid u)$ have bounded fourth order derivatives w.r.t. $x,u$ and $B$ for $B$ in a neighborhood of $B_0$.
	
	\item[(B4)] The matrix $M_0=E\left\{h^{(1)}(B_0^\T X)^\T h^{(1)}(B_0^\T X) \right\}$ has full rank $d$, where $h^{(1)}(\cdot)\in R^{s\times d}$ is the derivative matrix of $h(\cdot)$.
	
	\item[(B5)] $K(\cdot)$ is a symmetric univariate density function with bounded second order derivatives. All the moments of $K(\cdot)$ exist.
	
	\item[(B6)] Bandwidths $h_0=c_1n^{-r_h}$ where $0<r_h\leq 1/(p_0+6)$, $p_0=\max(p,3)$. For $t\geq 1$, $h_t=\max(r_nh_{t-1},h)$ where $r_n=n^{-r_h/2},h=c_2n^{-r_h'}$ with $0< r_h'\leq 1/(d+3)$, and $c_1,c_2$ are constants.
	\end{itemize}
	Define
	\begin{equation}\nonumber
		\begin{split} W_{B_0}&=E\left[\left\{h^{(1)}(B_0^\T X)^\T h^{(1)}(B_0^\T X)\right\}\otimes \left\{v_{B_0}(X)v_{B_0}^\T(X)\right\}\right],\\
			\Sigma_0&=\mathrm{var}\left[\left\{h^{(1)}(B_0^\T X)^\T\otimes v_{B_0}(X)\right\}\zeta   \right],\\
			W_0&=\mathrm{var}\left[\left\{M_0^{-1}h^{(1)}(B_0^\T X)^\T\zeta\right\}\otimes\left\{\bar{w}_{B_0}^+(X)v_{B_0}(X)\right\}\right].
		\end{split}
	\end{equation}
	
	\begin{theorem}\label{thm1}
		Under (A1)-(A6) and (B1)-(B6), the estimated $\hat{B}_{\mathrm{iMAVE}}$ from (\ref{eucmodel}) satisfies
		\begin{equation}\nonumber
			||\hat{B}_{\mathrm{iMAVE}}\hat{B}_{\mathrm{iMAVE}}^\T-B_0B_0^\T||_F=O(\h^3+\h\delta_{d\h}+\delta_{d\h}^2/h+n^{-1/2})
		\end{equation}
		in probability as $n\rightarrow\infty$, where $\delta_{d\h}=(n\h^d/\mlog n)^{-1/2}$. If $\h^3+\h\delta_{d\h}+\delta_{d\h}^2/h=o(n^{-1/2})$, then
		\begin{equation}\nonumber
			\sqrt{n}\left\{\vec(\hat{B}_{\mathrm{iMAVE}}\hat{B}_{\mathrm{iMAVE}}^\T B_0)-\vec(B_0)\right\}\stackrel{d}{\rightarrow}N(0,W_{B_0}^+\Sigma_0W_{B_0}^+).
		\end{equation}
	\end{theorem}
	
	Results in Theorem \ref{thm1} are consistent with those in Xia et al (2002), Xia (2007) and Zhang (2021). The iMAVE shares the merit of classic MAVE that it can achieve a faster consistency rate even without undersmoothing the nonparametric link function estimator. Similar results of iOPG are shown below.
	
	\begin{theorem}
		Under (A1)-(A6) and (B1)-(B6), the estimated $\hat{B}_{\mathrm{iOPG}}$ from (\ref{eucmodel}) satisfies
		\begin{equation}\nonumber
			||\hat{B}_{\mathrm{iOPG}}\hat{B}_{\mathrm{iOPG}}^\T-B_0B_0^\T||_F=O(\h^3+\h\delta_{d\h}+n^{-1/2})
		\end{equation}
		in probability as $n\rightarrow\infty$, where $\delta_{d\h}=(n\h^d/\mlog n)^{-1/2}$. If $\h^3+\h\delta_{d\h}=o(n^{-1/2})$, then
		\begin{equation}\nonumber
			\sqrt{n}\Big\{\vec(\hat{B}_{\mathrm{iOPG}}\hat{B}_{\mathrm{iOPG}}^\T B_0)-\vec(B_0)\Big\}\stackrel{d}{\rightarrow}N(0,W_0).
		\end{equation}
	\end{theorem}
	
	When $\sym$ is endowed with the log-Euclidean metric or the log-Cholesky metric, the manifold-related conditions are automatically satisfied and thus only (B1)-(B6) are needed. We present theoretical results of iMAVE and iOPG with $Y$ lying in $\sym$ endowed with the log-Euclidean metric in (\ref{eumave}) and (\ref{euopg}) by the following corollaries. The log-Cholesky case is almost the same and is omitted.
	
	In this case, $(\m,\oplus)$ with $\oplus$ defined in (\ref{groupop}) is an abelian Lie group and the bi-invariant log-Euclidean metric turns $\sym$ into a Hadamard manifold. Our model (\ref{ourmodel}) is valid and can be transformed into
	\begin{equation}\label{logmodel}
		\mlog Y=\mlog(g(B_0^\T X))+\mlog\varepsilon
	\end{equation}
	by the same reasoning in Section 5 (with $\mu$ replaced by $e$). We denote $h(B_0^\T X)=\mlog(g(B_0^\T X))$ and $\zeta=\mlog\varepsilon$. Terms in (\ref{logmodel}) are $m\times m$ symmetric matrices and if we vectorize the lower triangle part of these matrices into $m(m+1)/2$-dimensional vectors, then (\ref{logmodel}) coincides with (\ref{eucmodel}). Thus the main difference of the $\sym$ case is that $y_k,h_k,\zeta_k$ ($k=1,...,s$) in (B1)-(B6) should be replaced by $y_{kl},h_{kl},\zeta_{kl}$ ($1\leq l\leq k\leq m$) and $M_0$ in (B4) should be $M_{\mathrm{SPD}}=E\{\sum_{k=1}^{m}\sum_{l=1}^{k}h_{kl}^{(1)}(B_0^TX)h_{kl}^{(1)}(B_0^TX)^T \}$.
	
	Define
	\begin{equation}\nonumber
		\begin{split} W_{\mathrm{SPD}}&=E\left[\left\{\sum_{k=1}^{m}\sum_{l=1}^{k}h_{kl}^{(1)}(B_0^\T X)h_{kl}^{(1)}(B_0^\T X)^\T \right\}\otimes \left\{v_{B_0}(X)v_{B_0}^\T(X)\right\}\right],\\
			\Sigma_{\mathrm{SPD}}&=\mathrm{var}\left[\left\{\sum_{k=1}^{m}\sum_{l=1}^{k}h_{kl}^{(1)}(B_0^\T X)\zeta_{kl} \right\} \otimes v_{B_0}(X)  \right],\\
			W_0^{\mathrm{SPD}}&=\mathrm{var}\left[\left\{M_{\mathrm{SPD}}^{-1}\sum_{k=1}^{m}\sum_{l=1}^{k}h_{kl}^{(1)}(B_0^\T X)\zeta_{kl}\right\}\otimes\left\{\bar{w}_{B_0}^+(X)v_{B_0}(X)\right\}\right].
		\end{split}
	\end{equation}

	\begin{corollary}\label{coro1}
		Under(B1)-(B6),  the estimated $\hat{B}_{\mathrm{iMAVE}}$ from (\ref{eumave}) satisfies
		\begin{equation}\nonumber
			||\hat{B}_{\mathrm{iMAVE}}\hat{B}_{\mathrm{iMAVE}}^\T-B_0B_0^\T||_F=O(\h^3+\h\delta_{d\h}+\delta_{d\h}^2/h+n^{-1/2})
		\end{equation}
		in probability as $n\rightarrow\infty$, where $\delta_{d\h}=(n\h^d/\mlog n)^{-1/2}$. If $\h^3+\h\delta_{d\h}+\delta_{d\h}^2/h=o(n^{-1/2})$, then
		\begin{equation}\nonumber
			\sqrt{n}\left\{\vec(\hat{B}_{\mathrm{iMAVE}}\hat{B}_{\mathrm{iMAVE}}^\T B_0)-\vec(B_0)\right\}\stackrel{d}{\rightarrow}N(0,W_{\mathrm{SPD}}^+\Sigma_{\mathrm{SPD}}W_{\mathrm{SPD}}^+).
		\end{equation}
	\end{corollary}
	
	\begin{corollary}\label{coro2}
		Under(B1)-(B6),  the estimated $\hat{B}_{\mathrm{iOPG}}$ from (\ref{euopg}) satisfies
		\begin{equation}\nonumber
			||\hat{B}_{\mathrm{iOPG}}\hat{B}_{\mathrm{iOPG}}^\T-B_0B_0^\T||_F=O(\h^3+\h\delta_{d\h}+n^{-1/2})
		\end{equation}
		in probability as $n\rightarrow\infty$, where $\delta_{d\h}=(n\h^d/\mlog n)^{-1/2}$. If $\h^3+\h\delta_{d\h}=o(n^{-1/2})$, then
		\begin{equation}\nonumber
			\sqrt{n}\left\{\vec(\hat{B}_{\mathrm{iOPG}}\hat{B}_{\mathrm{iOPG}}^\T B_0)-\vec(B_0)\right\}\stackrel{d}{\rightarrow}N(0,W_0^{\mathrm{SPD}}).
		\end{equation}
	\end{corollary}
	
	In the proof of Corollary \ref{coro1} and Corollary \ref{coro2}, we would not encounter $\phi\log_{\hat{\mu}}Y_i-\log_\mu Y_i$. Actually even in the general manifold case $\phi\log_{\hat{\mu}}Y_i-\log_\mu Y_i$ does not have effects on the convergence rate and the asymptotic variance. As shown above, convergence rates in the general manifold case and the $\sym$ case are the same and asymptotic variances are consistent in form.
	
	\section{Determine the Structural Dimension}
	In this part, we discuss how to use a cross validation procedure to determine the structural dimension. We focus on the $\sym$ case and the method can be extended to general manifold similarly. Suppose $l$ is now the working dimension and $d$ is the true structural dimension. In the Euclidean case, Xia et al. (2002) defined
	\begin{equation}\nonumber
		\hat{a}_{l0,j}=\sum_{i=1,i\neq j}^nK_{h_l}^{(i,j)}y_i \Big/ \sum_{i=1,i\neq j}^nK_{h_l}^{(i,j)},
	\end{equation}
	where $y_i$ $(i=1,...,n)$ are scalars, $K_{h_l}^{(i,j)}=K_{h_l}(\hat{B}^\T(X_i-X_j))$ and the suffix $l$ is used to indicate that the bandwidth depends on the working dimension $l$. Actually $\hat{a}_{l0,j}$ is the N-W estimate of $y_j$. And the CV value is 
	\begin{equation}\nonumber
		\text{CV}(l)=\frac{1}{n}\sum_{j=1}^n(y_j-\hat{a}_{l0,j})^2\quad (l=1,...,p).
	\end{equation}
	
	In our case, $Y_i$ $(i=1,...,n)$ are now SPD matrices. If we equip $\sym$ with the log-Euclidean metric, then $\mlog Y_i$ $(i=1,...,n)$ are in $T_{I_m}\sym$. Similarly define
	\begin{equation}\nonumber
		\begin{split}
		\hat{a}_{l0,j}&=\sum_{i=1,i\neq j}^nK_{h_l}^{(i,j)}\vecs(\mlog Y_i) \Big/ \sum_{i=1,i\neq j}^nK_{h_l}^{(i,j)},\\
		\text{CV}(l)&=\frac{1}{n}\sum_{j=1}^n||\vecs(\mlog Y_i)-\hat{a}_{l0,j}||_F^2\quad (l=1,...,p).
		\end{split}
	\end{equation}
	where $||\cdot||_F$ is the matrix Frobenius norm. We then estimate $d$ as
	\begin{equation}\nonumber
		\hat{d}=\mathop{\arg}\min_{1\leq l\leq p}\text{CV}(l).
	\end{equation}
	
	\begin{theorem}\label{cv}
		Suppose assumptions (B1)-(B3) and (B5) hold. We have
		\begin{equation}\nonumber
			\lim_{n\rightarrow\infty}P(\hat{d}=d)=1.
		\end{equation}
	\end{theorem}
	
	Theorem \ref{cv} shows that as $n\rightarrow\infty$, the probability of choosing the right dimension tends to 1. If we equip $\sym$ with the log-Cholesky metric, above arguments still hold by replacing $\mlog Y_i$ with $\rm{chol}(Y_i)$.

\section{Simulation Studies}
\subsection{Study I for SPD Matrices }
In the following studies the structural dimension $d$ is known unless otherwise specified. 
We test the performance of our proposed iMAVE with log-Euclidean metric (eu-iMAVE), iOPG with log-Euclidean metric (eu-iOPG), iMAVE with log-Cholesky metric (ch-iMAVE), iOPG with log-Cholesky metric (ch-iOPG), weighted inverse regression ensemble method (WIRE, Ying and Yu (2022)), Fr\'echet MAVE and Fr\'echet OPG (fMAVE and fOPG, Zhang et al. (2021)). 

According to Schwartzman (2006), $Z\in \sy$ is said to obey the standard symmetric matrix variate Normal distribution $N_{mm}(0,I_m)$ if $Z$ has independent $N(0,1)$ diagonal elements and independent $N(0,1/2)$ off-diagonal elements. $Y\in\sy$ is said to obey the symmetric matrix variate Normal distribution $N_{mm}(M,\Sigma)$ if $Y=M+GZG^\T$ where $M\in\sy$ and $\Sigma=G^\T G$. As a special case, we say $Y\in\sy\sim N_{mm}(M,\sigma^2)$ if $Y=M+\sigma Z$.

Let $\beta_1^\T=(1,1,0,...,0)/\sqrt{2}$, $\beta_2^\T=(0,...0,1,1)/\sqrt{2}$. The predictors $X_1, X_2, ..., X_p$ are independent random variables each from the uniform distribution on $[0,1]$. We generate $n$ i.i.d samples $(X_{1i},X_{2i},...,X_{pi})$ $(i=1,...,n)$. Let $M(X)$ be matrices specified by the following models:


I-1: $M(X)=\left(
\begin{array}{cc}
	1 & \rho(X) \\
	\rho(X) & 1 
\end{array}\right)$, where $\rho(X)=\{\mexp(\beta_1^\T X)-1\}/\{\mexp(\beta_1^\T X)+1\}$;

I-2: $M(X)=\left(
\begin{array}{ccccc}
	1 & \rho_1(X) & \rho_1(X)&\rho_2(X) & \rho_2(X) \\
	\rho_1(X) & 1 & \rho_2(X)&\rho_2(X)&\rho_2(X) \\
	\rho_1(X) & \rho_2(X) & 1&\rho_2(X)&\rho_1(X)\\
	\rho_2(X)&\rho_2(X)&\rho_2(X)&1&\rho_1(X)\\
	\rho_2(X)&\rho_2(X)&\rho_1(X)&\rho_1(X)&1
\end{array}\right)$,\\
where  $\rho_1(X)=0.2\{\mexp(\beta_1^\T X)-1\}/\{\mexp(\beta_1^\T X)+1\}$ and $\rho_2(X)=0.2\sin(\beta_2^\T X)$.

We generate $\mlog(Y)\sim N_{mm}(\mlog\{M(X)\},\sigma^2)$. That is, $Y=\mexp[\mlog\{M(X)\}+\sigma Z]$. In model I-1, $m=2$, $B_0=\beta_1$ and $d=1$.  In model I-2, $m=5$, $B_0=(\beta_1,\beta_2)$ and $d=2$. In above settings $M(X)$ is not necessarily the Fr\'echet mean of $Y$ given $X$, but still measures the concentration tendency of the conditional distribution $Y\mid X$. Model I-1,I-2 are also considered in Zhang et al. (2021). The kernel function in iOPG and iMAVE is $K(v^2)=15/16(1-v^2)^2I(v^2<1)$. In WIRE, we adopt the distance function induced by the log-Euclidean metric to compute the distance matrix. We follow the same steps described in Zhang et al. (2021) to prepare  fOPG and fMAVE for the following simulations. For each model, we take $\sigma=0.2$ and $(p,n)=(10,100),(10,200),(20,100),(20,200)$. The experiments in each scenario was repeated 100 times and the means and standard deviations of the estimation errors are listed in Table \ref{table1}. The results for $\sigma=0.1$ are presented in the supplementary material.	
	
 \begin{table}[ht!]
\begin{center}
	\resizebox{0.8\columnwidth}{!}{
		\begin{tabular}{ccccccccc}
			\hline\\[-25pt]
			Model  & $(p,n)$ & WIRE & eu-iOPG & eu-iMAVE & ch-iOPG & ch-iMAVE & fOPG & fMAVE \\
			\hline\\[-25pt]
			\multirow{8}{*}{I-1}&(10,100)&0.0869&0.0693&0.0693&0.0616&0.0612&0.0891&0.3913\\
			~& &$\pm$0.0229&$\pm$0.0170&$\pm$0.0169&$\pm$0.0185&$\pm$0.0184&$\pm$0.0255&$\pm$0.2300\\
			~&(10,200)&0.0617&0.0489&0.0488&0.0421&0.0422&0.0592&0.3803\\
			~& &$\pm$0.0163&$\pm$0.0103&$\pm$0.0100&$\pm$0.0091&$\pm$0.0092&$\pm$0.0144&$\pm$0.2093\\
			~&(20,100)&0.1406&0.1118&0.1112&0.0973&0.0965&0.1443&0.3519\\
			~& &$\pm$0.0242&$\pm$0.0193&$\pm$0.0194&$\pm$0.0184&$\pm$0.0183&$\pm$0.0322&$\pm$0.1769\\
			~&(20,200)&0.0953&0.0735&0.0735&0.0656&0.0654&0.0934&0.2748\\
			~& &$\pm$0.0167&$\pm$0.0146&$\pm$0.0146&$\pm$0.0107&$\pm$0.0107&$\pm$0.0192&$\pm$0.1483\\
			\hline\\[-25pt]
			\multirow{8}{*}{I-2}&(10,100)&0.0577&0.0635&0.0605&0.0530&0.0532&0.2802&2.8665 \\
			~& &$\pm$0.0317&$\pm$0.0281&$\pm$0.0321&$\pm$0.0226&$\pm$0.0249&$\pm$0.4609&$\pm$0.2536\\
			~&(10,200)&0.0277&0.0283&0.0277&0.0246&0.0248&0.0504&2.9728\\
			~& &$\pm$0.0229&$\pm$0.0227&$\pm$0.0222&$\pm$0.0223&$\pm$0.0215&$\pm$0.0302&$\pm$0.0854\\
			~&(20,100)&0.1314&0.1656&0.1376&0.1192&0.1155&1.3775&3.0551\\
			~& &$\pm$0.0344&$\pm$0.0660&$\pm$0.0464&$\pm$0.0324&$\pm$0.0327&$\pm$0.5156&$\pm$0.3843\\
			~&(20,200)&0.0578&0.0582&0.0560&0.0525&0.0517&0.2223&2.9806\\
			~& &$\pm$0.0128&$\pm$0.183&$\pm$0.0144&$\pm$0.0148&$\pm$0.0136&$\pm$0.1651&$\pm$0.1680\\
			\hline
	\end{tabular}}
\end{center}
\setlength{\abovecaptionskip}{-10pt}
\captionsetup{font=footnotesize, labelfont=bf, labelsep=period}
\caption{Mean ($\pm$ standard deviation) of estimation error for different methods in study I.}
\label{table1}
\end{table}	

It is obvious that the best performer is always iOPG or iMAVE with either the log-Euclidean or the log-Cholesky metric. This result is reasonable since WIRE and fOPG, fMAVE make use of the information hidden in $Y$ by calculating the distance matrix  $(d(Y_i,Y_j))_{ij}$ or the kernel matrix $(k(Y_i,Y_j))_{ij}$, both of which  fail to fully exploit the inner structure of $Y$. On the contrary, our methods are intrinsic and respect the geometric structure of $Y$, thus generating more satisfying results.

\subsection{Study II for SPD Matrices}
In this simulation study, we generate $Y$ similar to Lin et al. (2022). Let the predictors $X_1,X_2,...,X_p$  be independently and identically sampled from the uniform distribution on $[0,1]$. Fix $\mu$ to be the identity matrix. Set $Y=\mu\oplus w(X_1,...,X_p)\oplus\zeta$,  where $w(X_1,...,X_p)=\mathfrak{exp}\phi_{\mu,e}f(X_1,...,X_p)$ with the following two settings for $f$:

II-1: $f(X_1,...,X_p)=f_{12}(X_1,X_2)$, where $f_{12}(X_1,X_2)$ is an $m\times m$ matrix with $(j,l)$-entry being $\mexp\{-1/|j-l|\}\sin[2\pi\{X_1+X_2-1/(j+l)\}]$;

II-2: $f(X_1,...,X_p)=\sum_{k=1}^2f_k(X_k)$ where $f_{k}(X_k)$ is an $m\times m$ matrix with $(j,l)$-entry being  $\mexp\{-1/|j-l|\}\sin[2\pi\{X_k-1/(j+l)\}]$.

The setting II-2 is the manifold additive model proposed by Lin  et al. (2022) and II-1 is a modification. We set $m=3$.  The random noise $\zeta$ is generated according to $\mathfrak{log}\zeta=\sum_{i=1}^6Z_jv_j$, where $Z_1,...,Z_6$ are independently sampled form $N(0,0.1^2)$ and $v_1,...,v_6$ are an basis of the tangent space $T_e\sym$. Note that $\mu$ is identical with $e$ so $\phi_{\mu,e}$ is just the identity map. We adopt the log-Euclidean metric so that $\mathfrak{exp}=\mexp$ and $\mathfrak{log}=\mlog$. In model II-1, $d=1$ and  $B_0=(1,1,0,...,0)^\T$; in model II-2, $d=2$ and $B_0=(\beta_1,\beta_2)^\T$, where $\beta_1=(1,0,...,0)^\T$ and $\beta_2=(0,1,0,...,0)^\T$. We take $(p,n)=(5,100),(5,200),(10,100),(10,200)$. Following Wang  et al. (2013) , we in this study adopt the multi-dimensional Gaussian kernel $k(u)=\mexp(-||u||^2/2)$ with the bandwidth $h$ set to be $h=\{4/(p+2)\}^{1/(p+4)}n^{-1/(d+4)}$ and $p$ being the dimension of $u$. The means and standard deviations of the estimation errors are summarized in Table \ref{table2}.

\begin{table}[ht!]
\begin{center}
	\resizebox{0.8\columnwidth}{!}{
		\begin{tabular}{ccccccccc}
			\hline\\[-25pt]
			Model &  $(p,n)$ & WIRE & eu-iOPG & eu-iMAVE & ch-iOPG & ch-iMAVE & fOPG & fMAVE \\
				\hline\\[-25pt]
			\multirow{8}*{II-1} & (5,100) & 1.2928&0.0872&0.0818&0.0871&0.0832&1.2666&1.2456\\
			~& & $\pm$0.1478&$\pm$0.2090&$\pm$0.2084&$\pm$0.2108&$\pm$0.2164&$\pm$0.1962&$\pm$0.2002\\
			~&(5,200)&1.2308&0.0280&0.0254&0.0291&0.0260&1.2124&1.2240\\
			~& & $\pm$0.1965&$\pm$0.0112&$\pm$0.0099&$\pm$0.0114&$\pm$0.0099&$\pm$0.2326&$\pm$0.2237\\
			~&(10,100)&1.3491&0.7189&0.6827&0.6925&0.6789&1.3413&1.3400\\
			~& &$\pm$0.0728&$\pm$0.6296&$\pm$0.6479&$\pm$0.6276&$\pm$0.6438&$\pm$0.0790&$\pm$0.0791\\
			~&(10,200)&1.3367&0.1641&0.1490&0.1500&0.1461&1.3320&1.3385\\
			~& &$\pm$0.1073&$\pm$0.3779&$\pm$0.3709&$\pm$0.3556&$\pm$0.3610&$\pm$0.1062&$\pm$0.0992\\
			\hline\\[-25pt]
			\multirow{8}*{II-2} &  (5,100) & 1.2118&0.0604&0.0554&0.0648&0.0594&1.2912&1.5003\\
			~& &$\pm$0.2560&$\pm$0.0195&$\pm$0.0169&$\pm$0.0195&$\pm$0.0176&$\pm$0.2745&$\pm$0.1855\\
			~&(5,200)&1.1923&0.0338&0.0331&0.0360&0.0352&1.2266&1.4979\\
			~& &$\pm$0.2578&$\pm$0.0093&$\pm$0.0092&$\pm$0.0099&$\pm$0.0099&$\pm$0.2354&$\pm$0.1562\\
			~&(10,100)&1.3954&0.3847&0.3651&0.3426&0.3246&1.6178&1.7211\\
			~& &$\pm$0.1070&$\pm$0.5302&$\pm$0.5309&$\pm$0.4995&$\pm$0.5047&$\pm$0.1532&$\pm$0.1396\\
			~&(10,200)&1.3714&0.0637&0.0566&0.0675&0.0603&1.4808&1.7123\\ 
			~& &$\pm$0.1005&$\pm$0.0121&$\pm$0.0104&$\pm$0.0126&$\pm$0.0108&$\pm$0.1526&$\pm$0.1334\\
				\hline
	\end{tabular}}
\end{center}
\setlength{\abovecaptionskip}{-10pt}
\captionsetup{font=footnotesize, labelfont=bf, labelsep=period}
\caption{Mean ($\pm$ standard deviation) of estimation error for different methods in study II.}
\label{table2}
\end{table}

\begin{figure}[ht!]
		\centering
		\includegraphics[width=0.7\linewidth]{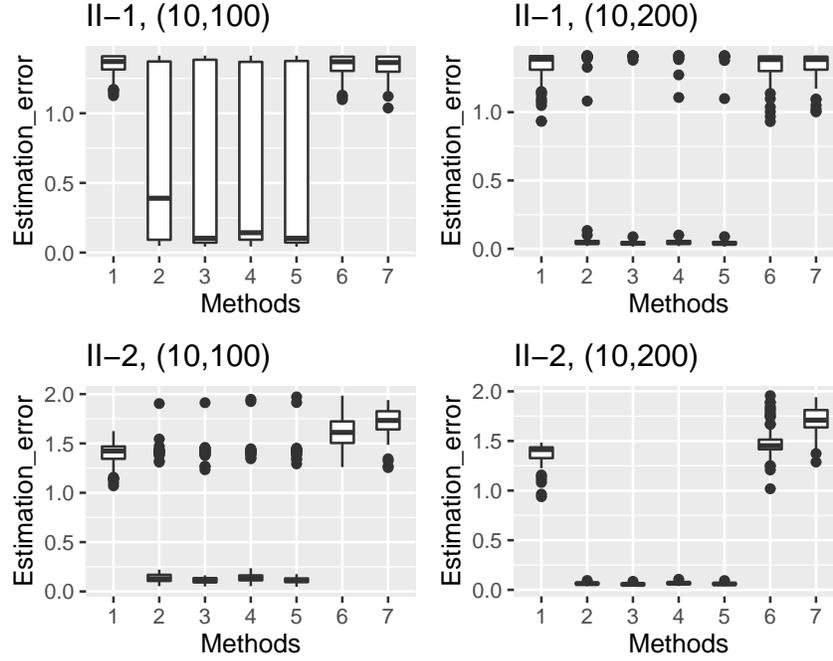}
		\setlength{\abovecaptionskip}{-1pt}
		\captionsetup{font=footnotesize, labelfont=bf, labelsep=period}
		\caption{Box plots of estimation errors in four scenarios. Numbers ``1-7" represent methods listed in Table \ref{table1} in turn.}
		\label{boxplot}
\end{figure}

Model II-1, II-2 are tough tasks, in each scenario all methods except ours fail to give reasonable estimates even when the dimension $p=5$ is not large at all. Our methods can give accurate estimates on most occasions. When the dimension is relatively large ($p=10$) and the sample size is not large enough ($n=100$), our  methods cannot always produce satisfying estimates and may fail. In Fig. \ref{boxplot}, we draw the box plots of the estimation errors based on 100 replications of all methods for $(p,n)=(10,100),(10,200)$ in II-1 and II-2. First we can see that WIRE, fOPG and fMAVE fail in all scenarios. When the sample size is not large enough ($p=100$), iOPG or iMAVE still has a possibility to fail even if the median of estimation errors is small and stable. See the top left box plot in Fig. \ref{boxplot}. The case II-2 is easier than II-1 for our models, with much less wrong estimates (the bottom two plots). When the sample size increases to 200, our methods improve themselves and give accurate estimates in every replication in II-2, while no obvious improvement is observed for other methods. It can be expected for our methods to produce more accurate estimates if the sample size is large enough.
	
\subsection{Study III for Sphere Data}
	Since the proposed iMAVE and iOPG can be extended to general manifolds, we in this part test the performance of models derived from model (\ref{eucmodel}). We generate $Y\in S^2$ according to the following model:
	
	III: Let $p_0=(0,0,1)^\T$ and the tangent vector at $p_0$ be
	\begin{equation}\nonumber
		l(X_i)=\left(\mathrm{exp}(X_{i1})\sin X_{i1}+\epsilon_{i1},\frac{\mathrm{exp}(X_{i1}+X_{i2})-1}{\mathrm{exp}(X_{i1}+X_{i2})+1}+\epsilon_{i2},0      \right)^\T.
	\end{equation}
	
	We generate i.i.d. observations $X_1,...,X_n$ from the uniform distribution on $[-1,1]$ and i.i.d. $\epsilon_{i1},\epsilon_{i2}\sim N(0,0.1^2)$. Then $Y_i$ is generated by
	\begin{equation}\nonumber
		Y_i=\mathrm{Exp}_{p_0}\{l(X_i)\}=\cos(||l(X_i)||)p_0+\sin(||l(X_i)||)l(X_i)/||l(X_i)||,
	\end{equation}
	where $||\cdot||$ is the Euclidean norm.
	
	The simulation results under several scenarios are listed in Table \ref{tablesphere}. The proposed iMAVE and iOPG always perform better than others, with iMAVE producing the smallest estimation errors.
	
	\begin{table}
	\begin{center}
		\resizebox{0.6\columnwidth}{!}{
			\begin{tabular}{ccccccc}
					\hline\\[-25pt]
				Model &  $(p,n)$ & WIRE & iOPG & iMAVE  & fOPG & fMAVE \\
					\hline\\[-25pt]
				\multirow{8}*{III} & (10,100)&0.3461&0.2555&0.2226&0.6743&1.5332\\
				~&&$\pm$0.0803&$\pm$0.0770&$\pm$0.0643&$\pm$0.2456&$\pm$0.1610\\
				~&(10,200)&0.2270&0.1545&0.1475&0.4065&1.5104\\
				~&&$\pm$0.0505&$\pm$0.0372&$\pm$0.0358&$\pm$0.1644&$\pm$0.0372\\
				~&(20,100)&0.5395&0.4766&0.3534&1.1215&1.6307\\
				~&&$\pm$0.1012&$\pm$0.0967&$\pm$0.0699&$\pm$0.2209&$\pm$0.1392\\
				~&(20,200)&0.3474&0.2481&0.2172&0.6990&1.6083\\
				~&&$\pm$0.0567&$\pm$0.0409&$\pm$0.0401&$\pm$0.1657&$\pm$0.1510\\
					\hline
		\end{tabular}}
	\end{center}
\setlength{\abovecaptionskip}{-10pt}
\captionsetup{font=footnotesize, labelfont=bf, labelsep=period}
\caption{Mean ($\pm$ standard deviation) of estimation error for different methods in study III.}
\label{tablesphere}
	\end{table}

\subsection{Study VI: Determine the Structural Dimension}
\begin{figure}
		\centering
		\includegraphics[width=0.9\linewidth]{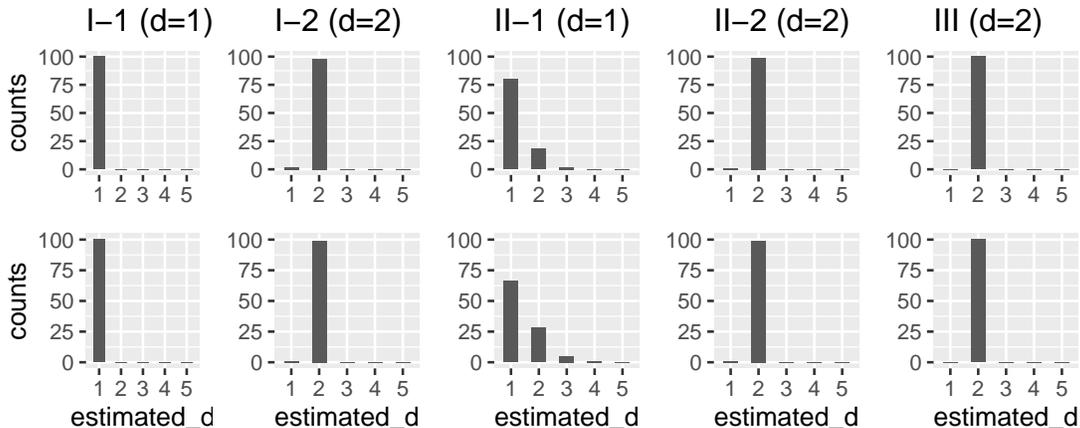}
		\setlength{\abovecaptionskip}{-1pt}
		\captionsetup{font=footnotesize, labelfont=bf, labelsep=period}
		\caption{Bar charts: counts of correct and false estimates in 100 times for five models with\\ $(p,n)=(10,200)$. $\sigma=0.1$ and $0.2$ correspond respectively the upper and the lower row.}
		\label{cvd}
	\end{figure}

	In this part we assume that we have no knowledge about the dimension of the mean dimension reduction space and need to estimate it. We generate data from the five models in Study I, II and III and use the CV procedure to estimate $d$. In the CV procedure, we use iOPG to estimate $B$. We set $p=10$, $n=200$ and repeat 100 times for each model and list the counts of correct and false estimates in 100 times when $\sigma=0.1$ and $0.2$, which is shown in Fig. \ref{cvd}.
	
	Except model II-1 with $\sigma=0.2$, our CV procedure  always gives satisfying estimations, reaching an accuracy greater than $80\%$ and even approaching $100\%$ in some cases. And if we increase the sample size to 300, the result corresponding to model II-1 with $\sigma=0.2$ becomes: $(\hat d<d):0$, $(\hat d=d):92$, $(\hat d>d):8$. Such improvement validates Theorem \ref{cv}.
\section{Application to New York Taxi Network Data}
In this section, we apply our proposed methods to the New York Taxi  network data. We first estimate the structural dimension as $\hat{d}$  and apply iMAVE equipped with the log-Euclidean metric to derive estimated $\hat{B}=(\hat{\beta}_1,...,\hat{\beta}_{\hat{d}})$ on the training dataset. Then we feed our results to the manifold additive regression model (Lin et al., 2022) and get the prediction root mean squared error (RMSE) on the testing dataset. Small RMSE will justify the validity of our methods.

The New York City Taxi and Limousine Commission provides records on pick-up and drop-off dates and times, pick-up and drop-off locations, trip distances, itemized fares, payment types and other information for yellow taxis (Tucker et al., 2021). The data are available from 

\texttt{https://www1.nyc.gov/site/tlc/about/tlc-trip-record-data.page}

Similar to Tucker et al. (2021), we transform raw data into network data (adjacent matrices), where zones  are nodes and edges are weighted by the number of taxi rides which picked up in one zone and dropped off in another within a single hour.  After proper mapping, these adjacent matrices lie in the space of SPD matrices. We do the following to collect SPD matrices together with several prediction variables:

1. We only choose the data of January and February, 2019 (59 days) due to resource restrictions.

2. We filter on observations with both pick-up and drop-off occurring in Manhattan (islands excluded).

3. We then group zones in Manhattan into 3 zones and label them similar to Dubey and Müller (2020). That is, each network has 3 nodes.

4. For each hour, we collected the number of pairwise connections between nodes based on pick-ups and drop-offs. These correspond to weights between nodes. We then further normalize the weights by the maximum edge weight in each hour so that they lie in $[0,1]$.

By doing so, we collected 1416 (59$\times$24) weighted adjacent matrices of $3\times 3$ describing the taxi movements between zones in Manhattan. To ensure that they are SPD matrices, we apply $\mexp(\cdot)$ to these symmetric matrices.

From the dataset we collect the following 9 potential predictors, with values averaged over each hour:

{\em Ave.Distance}: mean distance traveled, standardized

{\em Ave.Fare}: mean total fare, standardized

{\em Ave.Passengers}: mean number of passengers, standardized

{\em Ave.tip}: mean tip, standardized

{\em Cash}: sum of cash indicators for type of payment, standardized

{\em Credit}: sum of credit indicators for type of payment, standardized

{\em Dispute}: sum of  dispute indicators for type of payment, standardized

{\em Free}: sum of free indicators for type of payment, standardized

{\em LateHour}: indicator for the hour being between 11pm and 5am 

We also collect New York City weather data for January and February 2019 from

\texttt{https://www.wunderground.com/history/daily/us/ny/new-york-city/KLGA/date}

The following 5 weather variables are included as potential predictors:

{\em Ave.temp}: daily mean temperature, standardized

{\em Ave.humid}: daily mean humidity, standardized

{\em Ave.wind}: daily mean wind speed, standardized

{\em Ave.press}: daily mean barometric pressure, standardized

{\em Precip}: daily total precipitation, standardized

This then yields a total of 14 potential predictors. We can now write the data at hand as $\{Y,X_{n\times p}\}$, where $Y$ is an array of dimension $3\times3\times n$,  $n=1416$, $p=14$ and $Y[ , ,i]$ is a $3\times 3$ SPD matrix $(i=1,...,n)$. Then we randomly divide the dataset into a train set (991 samples) and a test dataset (425 samples). On the train set, we respectively set $d=1,...,7$, apply iMAVE with the log-Euclidean metric and calculate CV($d$). The results are: 0.0430, 0.0283, 0.0257, 0.0626, 0.0834, 0.0687, 0.0612. The CV procedure suggests that $\hat{d}=3$ is a reasonable choice. So we apply iMAVE with $d=3$ again to the training dataset and get $\hat{B}$ which is listed in Table \ref{bhat}.

\begin{table}
	\begin{center}
		\resizebox{1.0\columnwidth}{!}{
		\begin{tabular}{cccccccc}
			\hline\\[-25pt]
			{ Direction} &{ Ave.Distance}&{ Ave.Fare}&{ Ave.Passengers}&{ Ave.Tip}&{ Cash}&{ Credit}&{ Dispute}\\
			\hline\\[-25pt]
			$\beta_1$&0.2417&-0.4827&0.0927&-0.0313&-0.5720&0.5863&0.0074\\
			$\beta_2$&0.6592&-0.4002&0.2242&0.0878&0.5755&-0.0817&-0.0101\\
			$\beta_3$&-0.3931&-0.6700&-0.4348&0.0017&0.1952&-0.0343&-0.0577\\
			&{ Free}&{ LateHour}&{ Ave.Temp}&{ Ave.Humid}&{ Ave.Wind}&{ Ave.Press}&{ Precip}\\
			$\beta_1$&0.1277&0.0988&-0.0339&-0.0053&-0.0025&-0.2579&-0.0033\\
			$\beta_2$&-0.0174&0.0579&-0.0511&-0.0075&-0.0062&-0.0540&0.0064\\
			$\beta_3$&-0.1134&-0.3692&0.0789&-0.0030&-0.0163&0.0833&0.0470\\
			\hline
	\end{tabular}}
	\end{center}
\setlength{\abovecaptionskip}{-10pt}
\captionsetup{font=footnotesize, labelfont=bf, labelsep=period}
\caption{Estimated CS directions in New York taxi network data.}
\label{bhat}
\end{table}

The estimated results show that fare amount and type of payment are important covariates, which is consistent with the results of Tucker et al. (2021). Ave.Fare and Ave.Distance are closely related and both of them are significant in the first three directions. Cash and Credit are significant in the first direction, showing that most passengers tend to pay the fare by cash or credit. Another obvious observation is that all the 5 weather variables seem negligible since their coefficients are almost 0 in all of the first three directions. This is reasonable because as a global metropolitan, the New York City has established an advanced and robust public transportation system. And mild weather changes may have little compact on the function of taxi services. The weather condition during January and February 2019 is rather stationary, which accounts for the insignificance of weather variables.

To show our dimension reduction method is valid and has further statistical applications, we conduct the additive regression using the manifold additive model (MAM) introduced by Lin et al. (2022). The MAM is formulated as
\begin{equation}\nonumber
	Y=\mu\oplus w_1(X_1)\oplus...\oplus w_q(X_q)\oplus \zeta,
\end{equation}
where $Y$ is an SPD matrix, $\mu$ is the Fr\'echet mean of $Y$, each $w_k$ is function mapping $X_k$ into the SPD space, $\zeta$ is random noise which has a Fr\'echet mean corresponding to the group identity element, $X_i$ $(i=1,...,q)$ are scalar variables and $\oplus$ is the group operation.

We apply MAM to the train dataset after dimension reduction $\{Y^{\mathrm{train}},X^{\mathrm{train}}\hat{B}\}$ to get estimated $\hat{\mu}$ and functions $\hat{w}_1$, $\hat{w}_2$ and $\hat{w}_3$. Then we apply the trained MAM to the test dataset  $\{Y^{\mathrm{test}},X^{\mathrm{test}}\hat{B}\}$ to get the estimates $\hat{Y}^{\mathrm{test}}$. The prediction RMSE on the test dataset is 0.3220, which is a relative small number as the prediction error of a $3\times  3$ SPD. That is to say, MAM generates good estimation after processing data with our intrinsic dimension reduction method, which indicates that our method is valid and possesses the potential for widely applications.


	\bigskip
	\nocite{*}
	\bibliographystyle{apalike}
	
	\bibliography{mainbib.bib}

\begin{thebibliography}{}

\bibitem[Arsigny et~al., 2007]{Arsigny2007}
Arsigny, V., Fillard, P., Pennec, X., and Ayache, N. (2007).
\newblock Geometric means in a novel vector space structure on symmetric
  positive-definite matrices.
\newblock {\em {SIAM} Journal on Matrix Analysis and Applications},
  29:328--347.

\bibitem[Batchelor et~al., 2004]{Batchelor2004}
Batchelor, P.~G., Moakher, M., Atkinson, D., Calamante, F., and Connelly, A.
  (2004).
\newblock A rigorous framework for diffusion tensor calculus.
\newblock {\em Magnetic Resonance in Medicine}, 53:221--225.

\bibitem[Bhattacharjee and M\"{u}ller,
  2021]{https://doi.org/10.48550/arxiv.2108.05437}
Bhattacharjee, S. and M\"{u}ller, H.-G. (2021).
\newblock Single index {F}réchet regression. ar{X}iv:2108.05437 [stat.{ME}].

\bibitem[Chen et~al., 2020]{https://doi.org/10.48550/arxiv.2006.09660}
Chen, Y., Lin, Z., and M\"{u}ller, H.-G. (2020).
\newblock Wasserstein regression. ar{X}iv:2006.09660 [stat.{ME}].

\bibitem[Cook and Li, 2002]{Cook2002}
Cook, R.~D. and Li, B. (2002).
\newblock Dimension reduction for conditional mean in regression.
\newblock {\em The Annals of Statistics}, 30:455--474.

\bibitem[Cook and Weisberg, 1991]{Cook1991}
Cook, R.~D. and Weisberg, S. (1991).
\newblock Sliced inverse regression for dimension reduction: Comment.
\newblock {\em Journal of the American Statistical Association}, 86:328--332.

\bibitem[Cornea et~al., 2016]{Cornea2016}
Cornea, E., Zhu, H., Kim, P., and Ibrahim, J.~G. (2016).
\newblock Regression models on {R}iemannian symmetric spaces.
\newblock {\em Journal of the Royal Statistical Society: Series B (Statistical
  Methodology)}, 79:463--482.

\bibitem[Dubey and M\"{u}ller, 2020]{Dubey2020}
Dubey, P. and M\"{u}ller, H.-G. (2020).
\newblock Functional models for time-varying random objects.
\newblock {\em Journal of the Royal Statistical Society: Series B (Statistical
  Methodology)}, 82:275--327.

\bibitem[Fletcher et~al., 2004]{Fletcher2004}
Fletcher, P., Lu, C., Pizer, S., and Joshi, S. (2004).
\newblock Principal geodesic analysis for the study of nonlinear statistics of
  shape.
\newblock {\em {IEEE} Transactions on Medical Imaging}, 23:995--1005.

\bibitem[Kendall and Le, 2021]{Kendall2011}
Kendall, W.~S. and Le, H. (2021).
\newblock Limit theorems for empirical fréchet means of independent and
  non-identically distributed manifold-valued random variables.
\newblock {\em Brazilian Journal of Probability and Statistics},
  25(3):323--352.

\bibitem[Lang, 1999]{Lang1999}
Lang, S. (1999).
\newblock {\em Fundamentals of Differential Geometry}.
\newblock Springer New York.

\bibitem[Li, 2018]{Li2018}
Li, B. (2018).
\newblock {\em Sufficient Dimension Reduction}.
\newblock Chapman and Hall/{CRC}.

\bibitem[Li and Wang, 2007]{Li2007}
Li, B. and Wang, S. (2007).
\newblock On directional regression for dimension reduction.
\newblock {\em Journal of the American Statistical Association}, 102:997--1008.

\bibitem[Li, 1991]{Li1991}
Li, K.~C. (1991).
\newblock Sliced inverse regression for dimension reduction.
\newblock {\em Journal of the American Statistical Association}, 86:316--327.

\bibitem[Lin, 2019]{cholesky2019}
Lin, Z. (2019).
\newblock Riemannian geometry of symmetric positive definite matrices via
  {C}holesky decomposition.
\newblock {\em {SIAM} Journal on Matrix Analysis and Applications},
  40:1353--1370.

\bibitem[Lin and M\"{u}ller, 2021]{Linmuller2021}
Lin, Z. and M\"{u}ller, H.-G. (2021).
\newblock Total variation regularized fréchet regression for metric-space
  valued data.
\newblock {\em The Annals of Statistics}, 49(6):3510--3533.

\bibitem[Lin et~al., 2022]{Linetal2022}
Lin, Z., M\"{u}ller, H.-G., and Park, B.~U. (2022).
\newblock Additive models for symmetric positive-definite matrices and {L}ie
  groups.
\newblock {\em Biometrika}.

\bibitem[Lin and Yao, 2019]{LinandYao2019}
Lin, Z. and Yao, F. (2019).
\newblock Intrinsic riemannian functional data analysis.
\newblock {\em The Annals of Statistics}, 47:3533--3577.

\bibitem[Ma and Zhu, 2012]{Ma2012}
Ma, Y. and Zhu, L. (2012).
\newblock A semiparametric approach to dimension reduction.
\newblock {\em Journal of the American Statistical Association}, 107:168--179.

\bibitem[Ma and Zhu, 2013]{Ma2013}
Ma, Y. and Zhu, L. (2013).
\newblock Efficient estimation in sufficient dimension reduction.
\newblock {\em The Annals of Statistics}, 41:250--268.

\bibitem[Ma and Zhu, 2019]{Ma2019}
Ma, Y. and Zhu, L. (2019).
\newblock Semiparametric estimation and inference of variance function with
  large dimensional covariates.
\newblock {\em Statistica Sinica}, 29:567--588.

\bibitem[Pennec et~al., 2006]{Pennec2006}
Pennec, X., Fillard, P., and Ayache, N. (2006).
\newblock A {R}iemannian framework for tensor computing.
\newblock {\em International Journal of Computer Vision}, 66:41--66.

\bibitem[Petersen and M\"{u}ller, 2019]{Petersen2019}
Petersen, A. and M\"{u}ller, H.-G. (2019).
\newblock Fr{\'{e}}chet regression for random objects with {E}uclidean
  predictors.
\newblock {\em The Annals of Statistics}, 47:691--719.

\bibitem[Schwartzman, 2006]{Schwartzman2006}
Schwartzman, A. (2006).
\newblock Random ellipsoids and false discovery rates: statistics for diffusion
  tensor imagining data.
\newblock {\em {P}h{D} {T}hesis}, page Stanford University.

\bibitem[Terras, 1985]{Terras1985}
Terras, A. (1985).
\newblock {\em Harmonic Analysis on Symmetric Spaces and Applications I}.
\newblock Springer New York.

\bibitem[Tu, 2011]{Tu2011}
Tu, L.~W. (2011).
\newblock {\em An Introduction to Manifolds}.
\newblock Springer New York.

\bibitem[Tucker et~al., 2021]{Tucker2021}
Tucker, D.~C., Wu, Y., and M\"{u}ller, H.-G. (2021).
\newblock Variable selection for global {F}r{\'{e}}chet regression.
\newblock {\em Journal of the American Statistical Association}, pages 1--15.

\bibitem[Wang et~al., 2013]{Wang2013}
Wang, T., Xu, P., and Zhu, L. (2013).
\newblock Penalized minimum average variance estimation.
\newblock {\em Statistica Sinica}, 23:543--569.

\bibitem[Xia, 2006]{Xia2006}
Xia, Y. (2006).
\newblock Asymptotic distributions for two estimators of the single-index
  model.
\newblock {\em Econometric Theory}, 22:1112--1137.

\bibitem[Xia, 2007]{Xia2007}
Xia, Y. (2007).
\newblock A constructive approach to the estimation of dimension reduction
  directions.
\newblock {\em The Annals of Statistics}, 35:2654--2690.

\bibitem[Xia et~al., 2002]{Xia2002}
Xia, Y., Tong, H., Li, W.~K., and Zhu, L.-X. (2002).
\newblock An adaptive estimation of dimension reduction space.
\newblock {\em Journal of the Royal Statistical Society: Series B (Statistical
  Methodology)}, 64:363--410.

\bibitem[Ying and Yu, 2022]{Ying2022}
Ying, C. and Yu, Z. (2022).
\newblock Fr{\'{e}}chet sufficient dimension reduction for random objects.
\newblock {\em Biometrika}.

\bibitem[Yuan et~al., 2012]{Yuan2012}
Yuan, Y., Zhu, H., Lin, W., and Marron, J.~S. (2012).
\newblock Local polynomial regression for symmetric positive definite matrices.
\newblock {\em Journal of the Royal Statistical Society: Series B (Statistical
  Methodology)}, 74:697--719.

\bibitem[Zhang, 2021]{Zhang2021}
Zhang, H. (2021).
\newblock Minimum average variance estimation with group lasso for the
  multivariate response central mean subspace.
\newblock {\em Journal of Multivariate Analysis}, 184.

\bibitem[Zhang et~al., 2021]{https://doi.org/10.48550/arxiv.2110.00467}
Zhang, Q., Xue, L., and Li, B. (2021).
\newblock Dimension reduction and data visualization for {F}réchet regression.
  ar{X}iv:2110.00467 [stat.{ME}].

\bibitem[Zhu and M\"{u}ller, 2021]{https://doi.org/10.48550/arxiv.2105.05439}
Zhu, C. and M\"{u}ller, H.-G. (2021).
\newblock Autoregressive optimal transport models. ar{X}iv:2105.05439
  [stat.{ME}].

\bibitem[Zhu et~al., 2009]{Zhu2009}
Zhu, H., Chen, Y., Ibrahim, J.~G., Li, Y., Hall, C., and Lin, W. (2009).
\newblock Intrinsic regression models for positive-definite matrices with
  applications to diffusion tensor imaging.
\newblock {\em Journal of the American Statistical Association},
  104:1203--1212.

\end{thebibliography}
\end{document}